%% file: main.tex
\documentclass[preprint]{vgtc}

\graphicspath{{figures/}{pictures/}{images/}{./}} 

\usepackage{times}                     

\usepackage{tabu}                      
\usepackage{booktabs}                  
\usepackage{lipsum}                    
\usepackage{mwe}                       

\usepackage{mathptmx}                  

\usepackage{microtype}
\usepackage{multirow}
\usepackage{url}

\onlineid{0}

\vgtccategory{Research}

\vgtcinsertpkg

\newcommand{\bridge}{\emph{manvr3d}}

\providecolor{added}{rgb}{1,0,0}
\newcommand{\changed}[1]{#1}

\ieeedoi{10.1109/VIS60296.2025.00077}

\title{manvr3d: A Platform for Human-in-the-loop Cell Tracking in Virtual Reality}

\author{Samuel Pantze\thanks{e-mail: s.pantze@hzdr.de}\\
        \parbox{1.4in}{\scriptsize \centering CASUS, Görlitz, Germany\\
        Helmholtz-Zentrum Dresden-Rossendorf e.V.\\
        Dresden, Germany}
\and Jean-Yves Tinevez\thanks{e-mail: tinevez@pasteur.fr}\\
      \parbox{1.4in}{\scriptsize \centering Institut Pasteur\\Université Paris Cité\\Image Analysis Hub (IAH)\\75015 Paris, France}
\and Matthew McGinity\thanks{e-mail: matthew.mcginity@tu-dresden.de}\\
    \parbox{1.4in}{\scriptsize \centering IXLAB, Technische Universität Dresden, Germany}
\and Ulrik Günther\thanks{e-mail: ulrik.guenther@hzdr.de}\\
     \parbox{1.4in}{\scriptsize \centering Helmholtz-Zentrum Dresden-Rossendorf e.V.\\Dresden, Germany}
     }

\teaser{
  \centering
  \includegraphics[width=\linewidth]{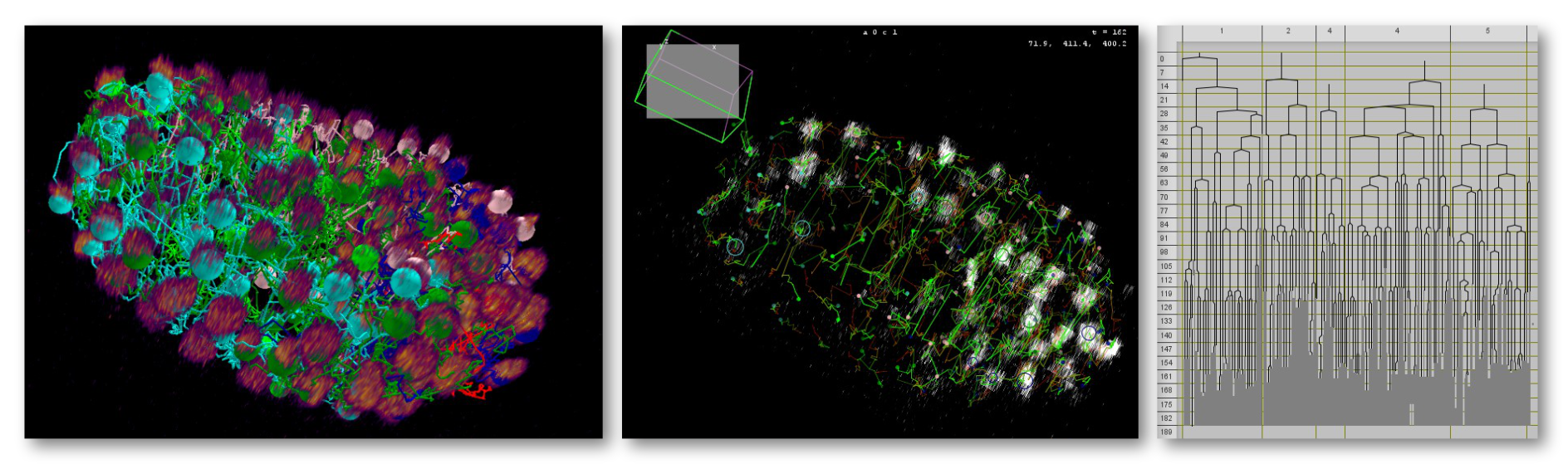}
  \caption{Cell tracking in Mastodon (middle), 3D visualization of volume and track data in sciview (left) and a cell lineage tree (right). The dataset shows the early development stage of a \emph{C. elegans} roundworm. Downloaded from the Cell Tracking Challenge website \cite{CTC}, data courtesy of Waterston Lab, University of Washington, Seattle, WA, USA \cite{elegans_dataset}.}
  \label{fig:teaser}
}

\abstract{
    \changed{We propose \bridge{}, a VR platform for immersive, AI-assisted human-in-the-loop cell tracking. Life scientists reconstruct the developmental history of organisms at the cellular level by analyzing 3D time-lapse microscopy images acquired at high spatio-temporal resolution. However, reconstruction of cell trajectories and lineage trees is a highly time consuming and error prone task. Common tools are often limited to 2D image display, which greatly limits spatial understanding and navigation. Deep Learning-based algorithms accelerate this process, yet depend heavily on manually-annotated, high-quality ground truth data and curation. In this work, we bridge the gap between Deep Learning-based cell tracking software and 3D/VR visualization to create a hybrid AI-human-in-the-loop cell tracking system. We lift the incremental annotation, training and proofreading loop of the deep learning model into the third dimension and apply natural user interfaces like hand gestures and eye tracking to accelerate the cell tracking workflow for life scientists. We present here the technical architecture of our platform and first analysis of performance. Our code is released open source.}
} 


\begin{document}



\maketitle

\section{Introduction} 

    Modern microscopes enable biologists to capture large scale 3D time-lapse datasets of embryonic development and other multi-cellular structures. Tracking of the imaged cells over time is a vital---yet non-trivial---task in the workflow of a life scientist studying the function and development of cells, tissues, and organisms. The result of the tracking process is a cell lineage tree (see \cref{fig:teaser}, right) that encodes information about the cellular ancestry.
    
    The tracking process consists classically of two stages.
    In the first \emph{detection} stage, cells are located in individual images or image stacks. In some cases, this might also include segmentation, in which cell shapes and boundaries are extracted in addition to cell positions. In the second \emph{linking} step, individual cells are matched between successive images, allowing cell trajectories and lineage to be extracted. 
    Tracking algorithms perform these tasks automatically on the input image. 
    This scientific topic has received considerable attention and many tools are available today. Recent algorithms rely either on conventional image processing, statistical methods (like Gaussian mixture models \cite{amat_fast_2014}) or deep learning methods \cite{deeplearningreview}. 
    Evaluating their performance or training supervised deep learning models require ground truth annotations, which are extremely time-consuming to create due to the effort involved in manually annotating cells and their links across frames. 
    
    One solution to this problem is to use sparse annotations combined with incremental human-in-the-loop deep learning. The neural cell-tracking model learns from human feedback, provided in the form of iterative cycles of corrections to the model's predictions. This is the approach taken by ELEPHANT \cite{elephant}. 
    However, with ELEPHANT, the human is constrained to a traditional mouse and keyboard interface and 2D display of 2D slices of the 3D data. This interface is not only slow but also error-prone, where the researcher might miss crucial spatial context.

    In this work, we present \bridge{}\ (Multimodal ANnotations in Virtual Reality 3D, pronounced \emph{``manfred''}), extending the approach taken by ELEPHANT by bringing the annotation and linking steps into virtual reality (VR) and enabling users to perform these tasks with VR controllers and eye tracking. The central contributions of this work are:

    \begin{enumerate}
        \item \bridge{}, a bridge between a widely-used, open-source cell tracking software and a 3D rendering engine, enabling bidirectional editing capabilities between 2D and 3D components, serving as a platform for developing natural user interface-based cell tracking solutions, and
        \item Two implementations of VR-based cell tracking---One using handheld controllers in an interactive VR environment, and another using eye-tracking hardware in VR to accelerate the tracking process even further.
    \end{enumerate}

    \changed{In this paper we describe the development of a functioning prototype. A full user study is beyond the scope of this work, in \cref{results} however, we show indicative numbers for both rendering and annotation performance.}

\section{Related work}

    Various software solutions exist for analyzing \changed{and visualizing} biological data in \changed{3D and/or VR environments}.
    \changed{For cell lineage data, visualization and annotation solutions exist for both 2D \cite{lange_aardvark_2025, pretorius_cell_2015} and 3D \cite{hong_lineaged_2022, salvador-martinez_celavi_2021}. Neither of these support immersive rendering in VR. Leeuw et al.  \cite{de_leeuw_visualization_2000} address this gap by rendering cell trajectories in a CAVE\footnote{Cave automatic virtual environment} system with superimposed volume time-series. None of those solutions integrate the visualization aspects into a wider analysis platform and do not offer integration of a machine learning model.
    
    \changed{Additionally, proprietary solutions like} \emph{ConfocalVR} \cite{stefani_confocalvr_2018} or \emph{syGlass} \cite{pidhorskyi_syglass_2018} offer general quantification and measurement tools \changed{for biological image data}, \emph{Arivis Pro VR}\footnote{\href{https://www.zeiss.com/microscopy/en/products/software/arivis-pro.html}{zeiss.com/microscopy/en/products/software/arivis-pro.html}} also offers features for cell tracking in volume data in VR. Arivis and syGlass have been utilized by Kaltenecker et al. \cite{kaltenecker_virtual_2024} to annotate cell segmentations, which they then use to train their 3D deep learning model \emph{DELiVR} for automated cell segmentation to speed up the annotation process. Note that in this work, no cell tracking is performed. Elor et al. developed the mixed reality environment \emph{BioLumin} \cite{elor_biolumin_2022} to study the efficacy of crowd-sourced tissue annotation tasks, where the collected data is later used for training deep learning models. VR has been used in \cite{guerinot_new_2022} to segment biological 3D time-series data, \cite{zhang_4d_2023} visualized meshes resulting from the segmentation process, and \cite{platt_micellanngelo_2023}  further annotate those meshes in VR. Again, no cell tracking is performed.

    In contrast to the aforementioned solutions, we present an immersive open-source visualization platform for human-in-the-loop cell tracking with natural user interfaces to enable a streamlined and ergonomic process for both the creation of ground truth data and the final proofreading step. Our system is embedded into the Fiji \cite{fiji} ecosystem, resulting in easier adoption and integration into existing pipelines.}

\section{Implementation} \label{method}

    We first detail the software ecosystem, then explain our platform together with the data structures used. Finally, we will describe both our handheld controller-based tracking solution, as well as the eye tracking-based one.

    \subsection{Software ecosystem}

        \bridge{} integrates with Fiji/ImageJ \cite{fiji}, a widely-used software package for both visualization and analysis of scientific and biological image data. For the purpose of this project, we rely on three existing Fiji plugins and frameworks:
        \begin{itemize}
            \item \emph{Mastodon} \cite{mastodon} as a platform for cell tracking, supporting the annotation of very large datasets. Mastodon uses \emph{BigDataViewer} \cite{pietzsch_bigdataviewer_2015} as a backend to efficiently render large volumetric data as 2D slices. Tracking is performed manually using mouse and keyboard interaction, or semi-automatically with a difference-of-Gaussians filter for cell detection and either a Kalman filter or LAP linker \cite{jaqaman_robust_2008} for cell linking.
            \item \emph{ELEPHANT} \cite{elephant} extends Mastodon with an incremental deep learning model. The model is first trained on a sparse dataset, consisting only of a handful of manual annotations. In a proofreading loop, the user then corrects the predictions and iteratively trains the model again and again, quickly increasing the size of the training dataset. A second (optional) U-Net model is trained on optical flow prediction, using the existing cell links as training data, to guide the linking of spots between time points.
            \item \emph{Sciview} \cite{sciview} and its underlying rendering framework \emph{scenery} \cite{scenery}, to allow visualization of large volumetric data together with 3D meshes. Both tools support a variety of natural user interfaces, such as VR headsets, or eye tracking hardware.
        \end{itemize}

        By extending widely-used open-source software packages, we aim to broaden the appeal of our platform to users, and encourage researchers to extend upon it further. Within Mastodon, \bridge{}\ acts as an extension and can be opened by selecting \texttt{Window $>$ New \bridge{}\ view}.
        

        

    \subsection{Software platform}

        \bridge{} is essentially a bridge that facilitates the interplay between Mastodon/ELEPHANT and sciview, enabling the reconstruction of a cell lineage tree in 3D, superimposed with a volume rendering of the image dataset. \bridge{}\ allows for bidirectional editing of the track data using VR controllers and other NUI devices. The same training and prediction commands that are available in the regular 2D interface of ELEPHANT are also available from within the VR environment. 



    \subsection{Organization and Data structure}\label{bridging-structures}

\begin{figure}[t]
            \centering
            \includegraphics[width=\linewidth]{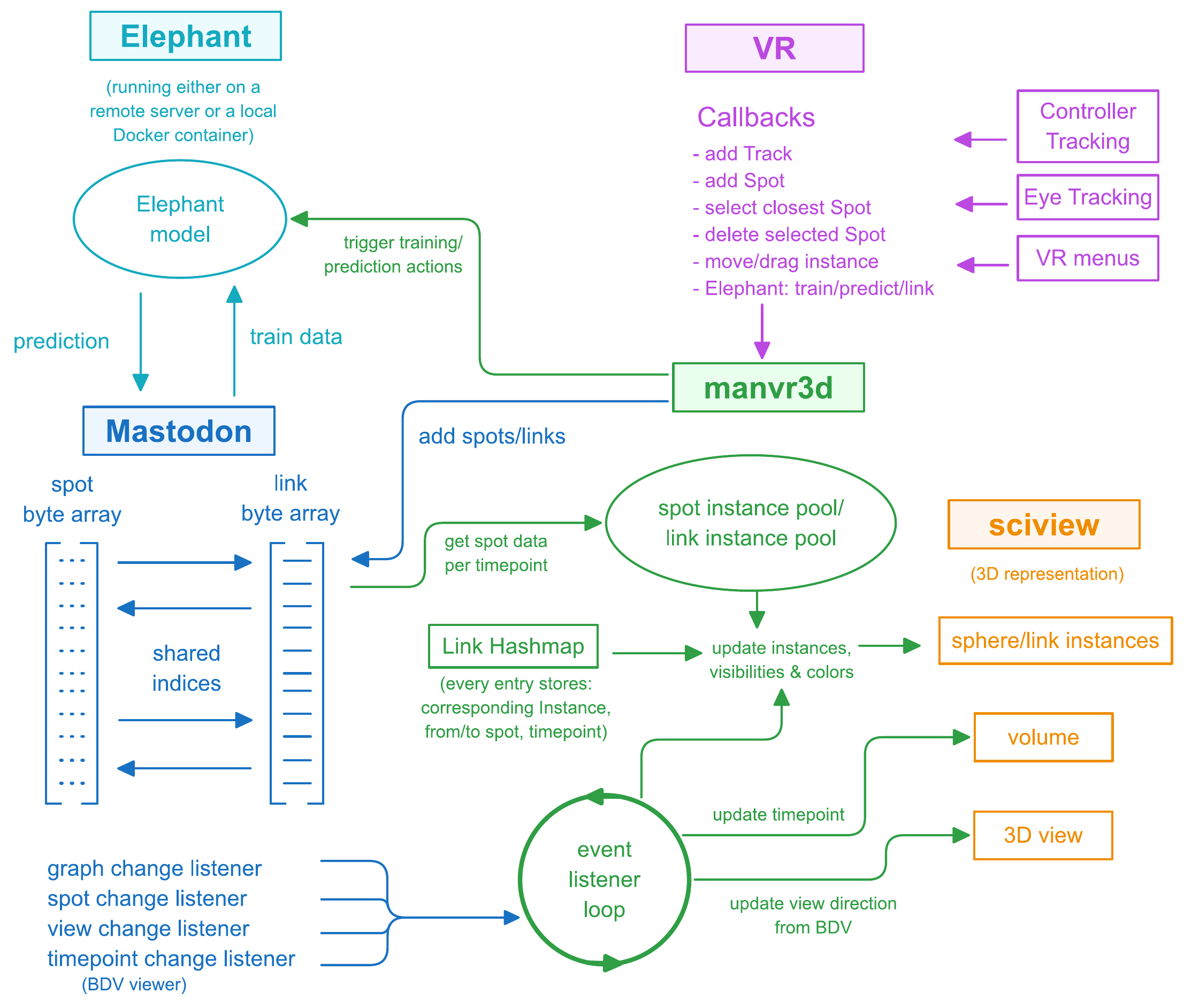}
            \caption{Data layout for \bridge{}. It maintains a 3D spot and track representation in sciview via an event listener loop. Editing events are handed back to Mastodon, where they trigger partial or full graph redraw events (see appendix for details).}
            \label{fig:data-layout}
        \end{figure}

        \bridge{}\ orchestrates all connections between the different components and ensures data consistency between the 2D data representation in Mastodon and the 3D representation in sciview. The data flow is visualized in \cref{fig:data-layout} with color coded components. \bridge{}\ (green) handles track editing events bidirectionally from either side and updates the other side accordingly. Time point changes are also communicated across components to maintain synchronized 2D (blue) and 3D (orange) views. Detailed information about Mastodons internal data structure are provided in Supplement \ref{appendix-datastructure}. We offer a graphical user interface that allows adjustment of various visualization parameters as well as launching a VR session (purple), either with or without additional hardware, such as eye trackers. The functionality of ELEPHANT (turquoise) is integrated as a menu inside the VR environment, enabling access to the training and prediction steps for the incremental deep learning model.

\begin{figure}[t]
            \centering
            \includegraphics[width=1.0\linewidth]{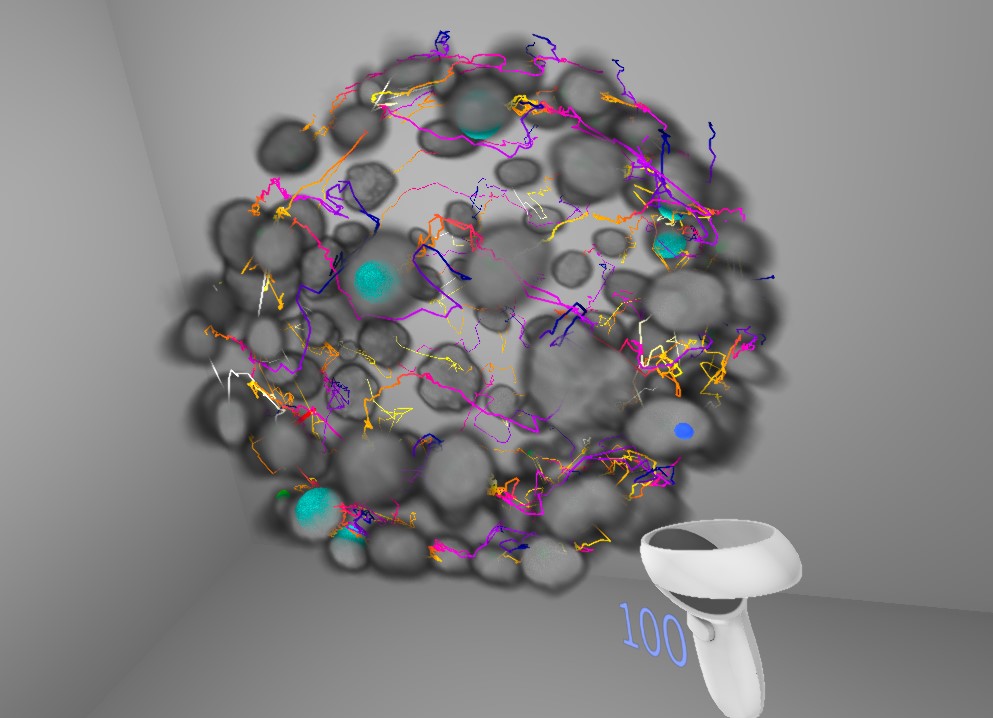}
            \caption{\changed{VR user looking at a \emph{Platynereis} dataset, annotated with controllers and ELEPHANT. Data courtesy of Tomancak Lab, MPI-CBG.}}
            \label{fig:controller-tracking}
        \end{figure}

        Annotated cell positions (spots) are only rendered for the current time point. Cell trajectories on the other hand span a longer time range, and as such they are treated independently of the spots. The bridge maintains a hash map of all links in the scene, which allows us to rapidly toggle the visibility of each segment to create a sliding window effect when moving through time. Using the time point information stored in the hash map, it is also possible to color each track with color maps that range from the minimum to the maximum time point. The effect of track coloring can be seen in \cref{fig:controller-tracking}. Individual spot editing events, like additions, movement and deletions, are handled on a per-spot and per-segment basis by event handlers and so do not cause a full graph redraw. To that end, the bridge locks update requests from event listeners on either side if an update is already in progress to prevent feedback loops and inconsistencies.
        
    

\section{Cell Tracking with Natural User Interfaces}

     We provide two implementations for tracking cells in VR with natural user interfaces: using controllers (and in the future possibly hand gestures) and an experimental implementation for utilizing eye tracking hardware, where the user's gaze directions are analyzed and cell tracks are created from this information. For intuitive interaction with the VR environment, we drew inspiration from popular VR productivity and creativity software like Gravity Sketch\footnote{\href{https://gravitysketch.com}{gravitysketch.com}} or Shapelab\footnote{\href{https://shapelabvr.com}{shapelabvr.com}}. Common one- and two-handed gestures for moving the observer, as well as scaling, rotating and translating of the dataset are implemented. A detailed controller layout can be found in Supplement \ref{appendix-interaction}. Clicking a spot highlights it, and allows the user to either move it to a new position or delete it. In the same manner, an arbitrary number of new spots can be placed to annotate all cells in the current time point.

     ELEPHANT actions are coupled to buttons on a wrist menu, allowing the user to interact with the deep learning model from within the VR environment. We currently support buttons for assigning the true positive label to all spots in the scene, for triggering the training, prediction and linking actions.
     
     These interaction methods act as a basis for both controller-based and eye tracking-based cell tracking.

    \subsection{Cell Tracking with Handheld Controllers}

        A 3D cursor in form of a small sphere is attached to the right VR controller and allows interaction with the VR environment. Cell position annotation is performed by moving the 3D cursor into the target cell and pressing the right trigger button. Through the semi-transparent rendering of the image data, it is possible to precisely position the cursor inside the desired cell. With each annotation click, time is automatically advanced, making the tracing of a cell very rapid, by repeatedly clicking into the cell. By default, time advances backwards, as this is found to simplify handling of cell division events. This process is repeated until either the first time point is reached or the user manually terminates the tracking process by pressing the right B button.

        The recorded track is sent to Mastodon only after tracking is finished to prevent continuous redrawing of the 3D representation. After the data are included in Mastodon's graph data structure, a full redraw event is triggered and the 3D tracks are updated.
        
        It is possible to merge an active track into an existing spot---a cell division, since the animation is played backwards---by simply clicking on the target spot. The opposite is also possible: Clicking an existing spot with the trigger button to start a new track will use that spot as its origin point. Using both of these features in conjunction thus allows for bridging holes in existing tracks.
        
        \begin{figure*}[h!]
            \centering
            \includegraphics[width=\linewidth, trim=0cm 0cm 0cm 0.23cm, clip]{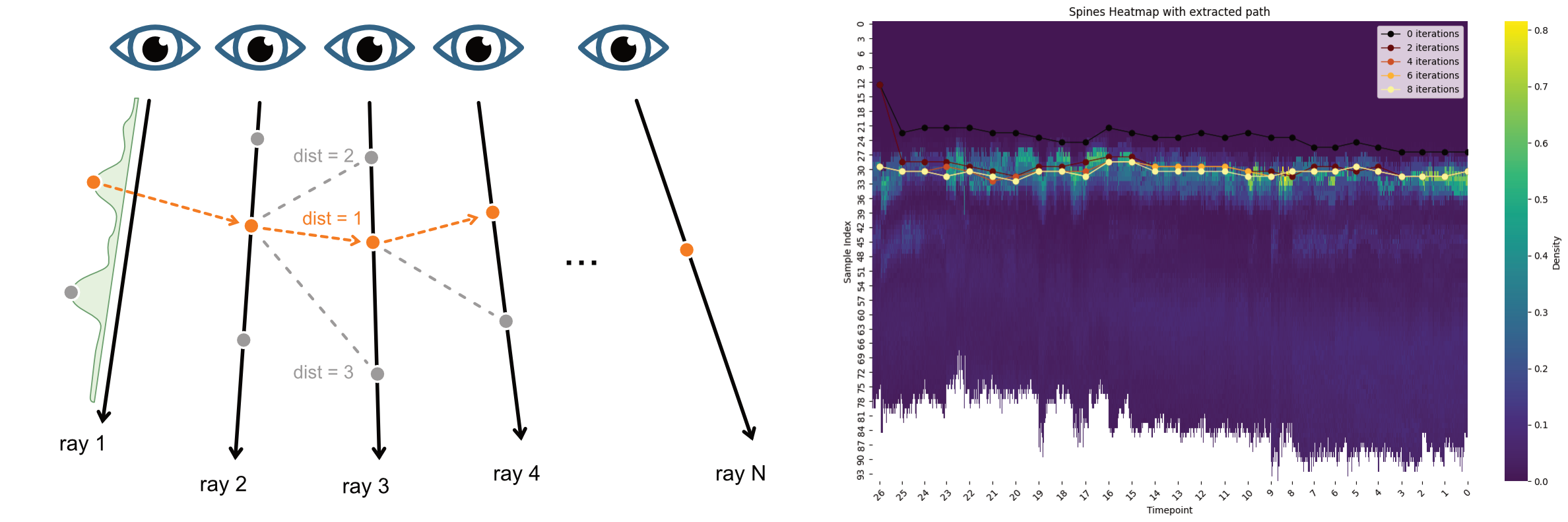}
            \caption{\changed{\emph{Left:} Scheme of the graph search algorithm that connects the closest local maxima in subsequent gaze rays following the first local maximum found in the first ray.}
            \emph{Right:} A collection of gaze rays across time, collected by following a moving cell with one's eyes. We slide a simple Gauss kernel in the shape of $[0.25, 0.5, 0.25]$ along each ray to smooth the signal and extract local maxima more effectively. Here we show the effect of various iterations of Gauss smoothing. It can be seen that no or low amounts of smoothing can lead to incorrectly extracted tracks, and starting from 4 iterations onward we are able to extract the correct cell track.}
            \label{fig:graph-spines-combined}
        \end{figure*}
        
    \subsection{Cell Tracking via Eye Tracking}

        Tracking cells using gaze interaction has been explored in \emph{Bionic Tracking} \cite{gunther_bionic_2020} by Günther et al. We incorporate this technique into our platform. While following a moving cell through the 3D volume data with their gaze, we record the user's gaze directions and sample the volume at a uniform interval along each gaze ray. \changed{Finding the position of a cell along this gaze ray thus turns into a 1-dimensional problem, because we can assume that the first local maximum along the ray corresponds to the target cell. After calculating the local maxima for each ray, the track is reconstructed with a variant of the A* algorithm \cite{sun_dynamic_2009} that connects the closest local maxima from subsequent gaze rays, see \cref{fig:graph-spines-combined}.}

        To avoid the Midas touch problem \cite{jacob_eye_1995}, it is important to remove any visual distractions that could lead the user to unintentionally look away from their target cell. For this reason, we implemented a dual input approach for cell tracking by eye tracking: as soon as the user is ready to start tracking a cell with their eyes, they press the left trigger button. This starts playback of the dataset and the continuous collection of gaze directions and volume density samples along each ray. Once the user interrupts the tracking with the trigger button, or if the first time point is reached, the collected gaze rays and their sampled values are analyzed and subsequently sent to Mastodon. A collection of rays is plotted in \cref{fig:graph-spines-combined,}, where each ray originates from the positive Y axis and extends downwards. Changes over time are plotted along the X axis. 


\section{\changed{Performance and Results}}\label{results}
\subsection{\changed{Rendering Performance}}

        In \bridge{}, we reuse the same underlying data allocated by the BigDataViewer backend in Mastodon. Especially with large datasets, this reduces memory load significantly, compared to solutions where memory sharing is not possible and copies are necessary.

        Both primitive types of the 3D representation, spheres for spots and cylinders for links, are currently rendered as instanced meshes on the GPU. Two instance pools are populated during the initialization phase---one for spots and one for links. We found that instance pre-generation is faster than on-the-fly generation. The instances in these pools are then positioned and colored according to the current time point. If a time point requires more spots, they are dynamically allocated to the pool.

        We benchmarked \bridge{}'s capability to render geometry instances (see \cref{tab:performance}) with two differently-sized datasets and found that for up to several ten thousand cells, there is no negative effect on performance\footnote{The frame rates refer to a render resolution of 1920x1080 pixels, on an XMG Fusion 15 laptop with NVIDIA RTX 4070 GPU and Intel Core i9-14900HX processor, running Windows 10 build 19044.5371, JDK version 21.0.5, NVIDIA driver version 566.36.}. Turning off the overlaid volume rendering yields slightly faster frame rates.
        
        \begin{table}[h]
            \centering
            \caption{Frame rate and scene population time, depending on the number of geometry instances rendered in the scene. The scene population time is only required once during scene initialization.}
            \begin{tabular}{@{}lcc@{}}
                \toprule
                Dataset size & Small & Large \\ \midrule
                \multicolumn{1}{l|}{Number of links} & 3000 & 243,000 \\
                \multicolumn{1}{l|}{Spots rendered per time point} & 90-110 & 2500-3700 \\
                \multicolumn{1}{l|}{Frame rate with volume (fps)} & 185 & 29 \\
                \multicolumn{1}{l|}{Frame rate without volume (fps)} & 230 & 32 \\
                \multicolumn{1}{l|}{Scene population time (s)} & 0.2 & 15 \\ \bottomrule
                \end{tabular}
            
            \label{tab:performance}
        \end{table}

\vspace{-1em}
\subsection{\changed{Annotation performance}}
\label{subsec:annotation-performance}
    \changed{
    As stated before, performing a full user study is beyond the scope of this work. We instead performed two partial annotations (\cref{tab:annotation-speed}) of a \emph{Platynereis} dataset and a \emph{Drosophila} dataset \cite{drosophila-dataset}, to compare annotation speeds between the Mastodon approach (2D), VR-based controller tracking (VR) and ELEPHANT (+DL). We measured the time taken to create 10 tracks each. We then used the pre-trained \emph{versatile} ELEPHANT model and trained it over 5 epochs on the annotations created, and divided the final amount of tracks (104 for \emph{Platynereis} and 74 for \emph{Drosophila}) by the sum of training and prediction time. The training/prediction time is independent of the manual annotation method used.

    \begin{table}[h]
    \centering
    \caption{Annotation times for 2D, VR and Deep Learning tracking.}
    \label{tab:annotation-speed}
    \begin{tabular}{@{}llll@{}}
    \toprule
    Dataset                      & Size                                  & Type & Time/track \\ \midrule
    \multirow{2}{*}{\emph{Platynereis}} & \multirow{2}{*}{700x660x113, 101 time points} & 2D   & 3.85 min   \\
                                 &                                       & \textbf{VR}   & \textbf{0.65 min}   \\
                                 &                                       & +DL   & 0.125 min  \\
    \emph{Drosophila}                   & 151x101x29, 31 time points                    & 2D   & 1.07 min   \\
                                 &                                       & \textbf{VR}   &
                                 \textbf{0.16 min}   \\
                                 &                                       & +DL   & 0.02 min   \\ \bottomrule
    \end{tabular}
    \end{table}
    
    These numbers indicate that VR-based tracking can significantly outperform manual 2D annotations, by a factor of about 6. Although the time needed for prediction and training in the ELEPHANT model remains the same for both methods, \bridge{} can provide a significant benefit for the annotation aspect.
    }

\section{Summary and future work} \label{summary}

    In this work, we presented \bridge{}, a platform to allow VR/natural user interface-based cell tracking together with two example implementations, using handheld controllers or eye tracking hardware. We couple this annotation process with the ELEPHANT deep learning model for rapid training data acquisition and provide track editing tools for proofreading of the predictions. \bridge{} easily scales to several thousand cells.
    
    The ELEPHANT model provides uncertainty information to indicate the confidence of the network's prediction. We plan to incorporate these data into the visualization process by coloring the spots and tracks accordingly, thus guiding the user towards areas of potentially higher error rate. Visualizing these uncertainty data is still an area of active research \cite{uncertainty}.
    
    \changed{To quantify the improvement---indicated in \cref{subsec:annotation-performance}---of our human-in-the-loop tracking approach over conventional methods, we plan to conduct a user study that compares the speed and accuracy of fully manual methods with automated methods and the approach taken in this project, using a variety of datasets.}

\section{Software Availability}
    The software can be found in the Github repository at \href{https://github.com/scenerygraphics/manvr3d}{github.com/scenerygraphics/manvr3d}. A fully-packaged version for easy deployment on Windows systems can be downloaded at \href{https://github.com/scenerygraphics/manvr3d/releases/}{github.com/scenerygraphics/manvr3d/releases/}.





\acknowledgments{The authors thank Ko Sugawara for his ongoing support with the ELEPHANT integration. Thanks to Vladimír Ulman and Ruoshan Lan for their work on a prototype of this project, and to Jan Tiemann for providing code and support for 3D user interface elements. The authors also want to thank Elias Barriga, Valentin Ruffine, Ana Patricia Ramos and Jaime Hidalgo for their feedback on the software.

SP was partially funded by the Center for Advanced Systems Understanding (CASUS), financed by Germany's Federal Ministry for Research, Technology, and Space (BMFTR) and by the Saxon Ministry for Science, Culture and Tourism (SMWK) with tax funds on the basis of the budget approved by the Saxon State Parliament. JYT was funded by the French National Research Agency (France BioImaging, ANR-24-INBS-0005 FBI BIOGEN). UG was supported by \emph{GoBio Initial}, 03LW0622, by the Germany's Federal Ministry for Research, Technology, and Space (BMFTR).
}
\bibliographystyle{abbrv-doi-hyperref-narrow}

\bibliography{main}

\appendix 

\input{appendix}

\end{document}

%% file: appendix.tex
\clearpage
\section{Supplementary Material}


    \begin{figure*}[t]
            \centering
            \includegraphics[trim=0cm 0cm 0cm 2cm, clip, width=\linewidth]{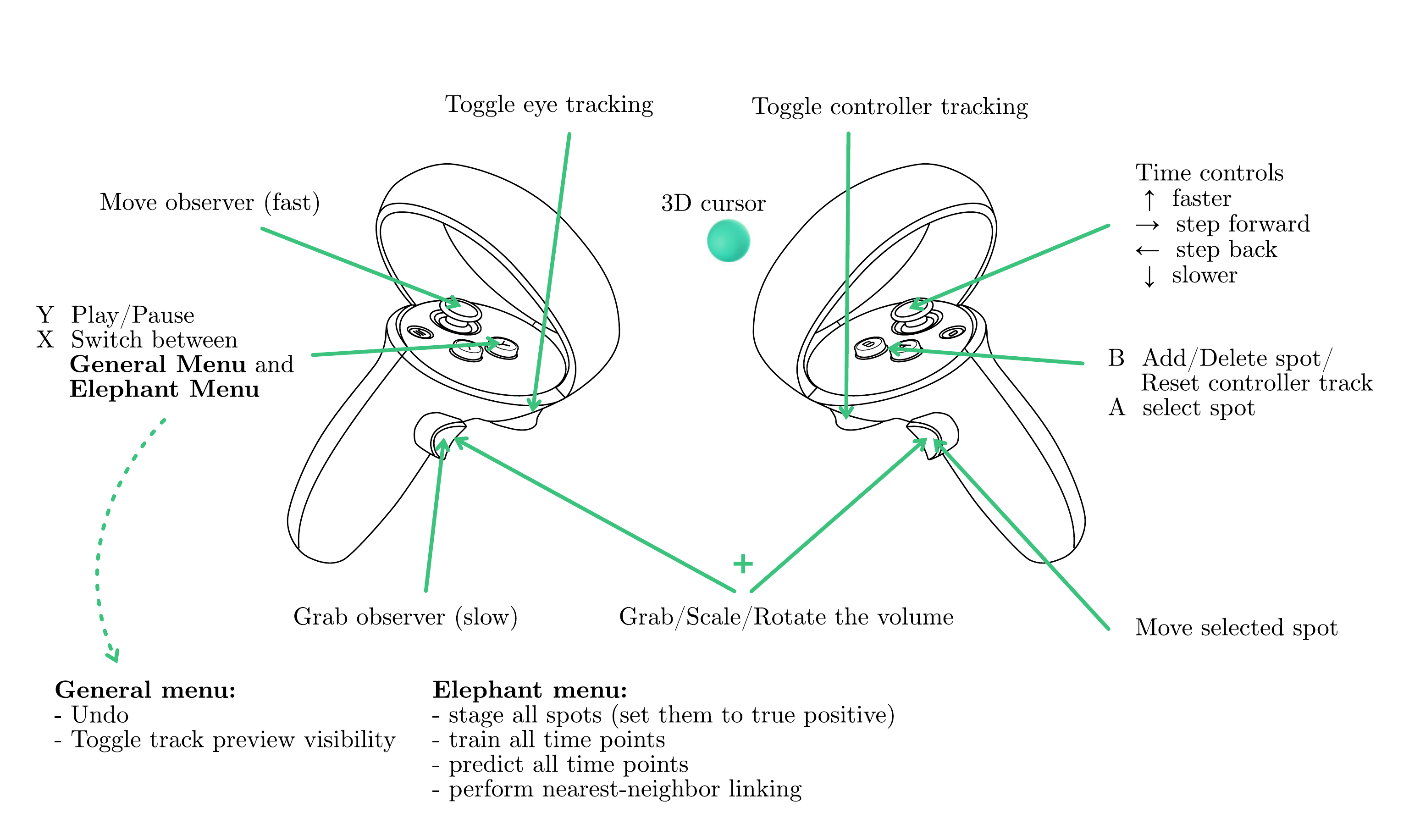}
            \caption{VR controller layout for a pair of Meta Quest 2 controllers.}
            \label{fig:controller-layout}
        \end{figure*}

\subsection{Mastodon data structure}\label{appendix-datastructure}

    The data structure employed by Mastodon is highly efficient, with the entire directed graph stored internally as primitive byte arrays \cite{mastodon}. The reasons for this are vastly improved data access speed and memory efficiency compared to storing all vertices and edges of the graph as Java objects. One array stores all spots -- the cell positions -- and a second array stores all links that connect those spots. Both arrays reference each other using indices, and their content is accessed via proxy objects and iterators. A single spot stores data for the first incoming and outgoing link, respectively, as well as spot attributes like color and a covariance matrix that contains the spot’s scale and rotation. In addition to the indices for its source and target spots, each link also stores the indices for the next source or target link in the case of merge or split events.
    
    Spot colors can be customized in Mastodon using tag sets; this aids visual analysis of the dataset. \bridge{} supports transferring these colors to the 3D tracks.

\subsection{VR interactions}\label{appendix-interaction}

    The VR controller layout is currently optimized for a Meta Quest 2, as it is an affordable yet capable VR headset and shares its layout with many other models and brands. Different controller layouts will be supported at a later time.

    Eye tracking and controller-based tracking interactions are paired to the left and right trigger buttons. The left grab button moves the observer through the scene. Fast movements are possible with the left joystick. Pressing both grab buttons will scale, rotate and translate the dataset.

    The left X button cycles between two wrist menus with buttons. One menu comprises an undo function that is coupled to Mastodon's undo recorder, as well as a toggle for preview track visibility during controller-based tracking. The second menu offers ELEPHANT actions. The first command prepares all spots in the scene for model training by assigning them the \emph{true positive} label. The other commands trigger the training, prediction and linking actions, respectively.
    
    Time controls are implemented via the left Y button for play/pause functionality, and the right joystick to move through the timeline and to change the speed of automatic playback.

    The user can highlight existing spots via the 3D cursor by selecting them with the right A button. A selected spot can then be deleted with the right B button. If no spot is selected, the B button will add a new spot to the scene at the current cursor position instead. Selected spots can be repositioned by holding the right grab button.

%% file: main.bib
@misc{drosophila-dataset,
  author       = {William Lemon},
  title        = {Drosophila embryo tissue time-lapse.},
  month        = jul,
  year         = 2019,
  publisher    = {Zenodo},
  version      = {1.0.1},
  doi          = {10.5281/zenodo.3336346},
  url          = {https://doi.org/10.5281/zenodo.3336346}
}

@INPROCEEDINGS{scenery,
  author={Günther, Ulrik and Pietzsch, Tobias and Gupta, Aryaman and Harrington, Kyle I.S. and Tomancak, Pavel and Gumhold, Stefan and Sbalzarini, Ivo F.},
  booktitle={2019 IEEE Visualization Conference (VIS)}, 
  title={scenery: Flexible Virtual Reality Visualization on the Java VM}, 
  year={2019},
  volume={},
  number={},
  pages={1-5},
  doi={10.1109/VISUAL.2019.8933605}
}

@inproceedings{sciview,
booktitle = {VisGap - The Gap between Visualization Research and Visualization Software},
editor = {Gillmann, Christina and Krone, Michael and Reina, Guido and Wischgoll, Thomas},
title = {{Tales from the Trenches: Developing sciview, a new 3D viewer for the ImageJ community}},
author = {Günther, Ulrik and 
Harrington, Kyle I. S.},
year = {2020},
publisher = {The Eurographics Association},
ISBN = {978-3-03868-125-0},
DOI = {10.2312/visgap.20201112}
}

@article{fiji,
author = {Schindelin, Johannes and Arganda-Carreras, Ignacio and Frise, Erwin and Kaynig, Verena and Longair, Mark and Pietzsch, Tobias and Preibisch, Stephan and Rueden, Curtis and Saalfeld, Stephan and Schmid, Benjamin and Tinevez, Jean-Yves and White, Daniel and Hartenstein, Volker and Eliceiri, Kevin and Tomancak, Pavel and Cardona, Albert},
year = {2012},
month = {06},
pages = {676-82},
title = {Fiji: An Open-Source Platform for Biological-Image Analysis},
volume = {9},
journal = {Nature methods},
doi = {10.1038/nmeth.2019}
}

@misc{mastodon,
    author = {Pietzsch, Tobias and Tinevez, Jean-Yves and Arzt, Matthias and Ulman, Vladimír and Sugawara, Ko and Hahmann, Stefan},
    note = {https://mastodon.readthedocs.io/en/latest/},
    title = "Mastodon"
}

@article {elephant,
    author = {Sugawara, Ko and {\c{C}}evrim, {\c{C}}a?r? and Averof, Michalis},
    title = {Tracking cell lineages in 3D by incremental deep learning},
    year = {2022},
    doi = {10.7554/eLife.69380},
    URL = {https://doi.org/10.7554/eLife.69380},
    journal = {eLife}
}

@article{uncertainty,
title = {A review of uncertainty quantification in deep learning: Techniques, applications and challenges},
journal = {Information Fusion},
volume = {76},
pages = {243-297},
year = {2021},
issn = {1566-2535},
doi = {https://doi.org/10.1016/j.inffus.2021.05.008},
url = {https://www.sciencedirect.com/science/article/pii/S1566253521001081},
author = {Moloud Abdar and Farhad Pourpanah and Sadiq Hussain and Dana Rezazadegan and Li Liu and Mohammad Ghavamzadeh and Paul Fieguth and Xiaochun Cao and Abbas Khosravi and U. Rajendra Acharya and Vladimir Makarenkov and Saeid Nahavandi}
}

@article{deeplearningreview,
	title = {Deep learning for cellular image analysis},
	volume = {16},
	issn = {1548-7091, 1548-7105},
	url = {http://www.nature.com/articles/s41592-019-0403-1},
	doi = {10.1038/s41592-019-0403-1},
	language = {en},
	number = {12},
	urldate = {2020-08-15},
	journal = {Nature Methods},
	author = {Moen, Erick and Bannon, Dylan and Kudo, Takamasa and Graf, William and Covert, Markus and Van Valen, David},
	month = dec,
	year = {2019},
	pages = {1233--1246},
}

@article{elor_biolumin_2022,
	title = {{BioLumin}: {An} {Immersive} {Mixed} {Reality} {Experience} for {Interactive} {Microscopic} {Visualization} and {Biomedical} {Research} {Annotation}},
	volume = {3},
	shorttitle = {{BioLumin}},
	url = {https://dl.acm.org/doi/10.1145/3548777},
	doi = {10.1145/3548777},
	number = {4},
	urldate = {2024-02-12},
	journal = {ACM Transactions on Computing for Healthcare},
	author = {Elor, Aviv and Whittaker, Steve and Kurniawan, Sri and Michael, Sam},
	month = nov,
	year = {2022},
	pages = {44:1--44:28},
}

@article{kaltenecker_virtual_2024,
	title = {Virtual reality-empowered deep-learning analysis of brain cells},
	copyright = {2024 The Author(s)},
	issn = {1548-7105},
	url = {https://www.nature.com/articles/s41592-024-02245-2},
	doi = {10.1038/s41592-024-02245-2},
	language = {en},
	urldate = {2024-04-26},
	journal = {Nature Methods},
	author = {Kaltenecker, Doris and Al-Maskari, Rami and Negwer, Moritz and Hoeher, Luciano and Kofler, Florian and Zhao, Shan and Todorov, Mihail and Rong, Zhouyi and Paetzold, Johannes Christian and Wiestler, Benedikt and Piraud, Marie and Rueckert, Daniel and Geppert, Julia and Morigny, Pauline and Rohm, Maria and Menze, Bjoern H. and Herzig, Stephan and Berriel Diaz, Mauricio and Ertürk, Ali},
	month = apr,
	year = {2024},
	note = {Publisher: Nature Publishing Group},
	pages = {1--10},
}

@article{stefani_confocalvr_2018,
	title = {{ConfocalVR}: {Immersive} {Visualization} for {Confocal} {Microscopy}},
	volume = {430},
	issn = {0022-2836},
	shorttitle = {{ConfocalVR}},
	url = {https://www.sciencedirect.com/science/article/pii/S0022283618306648},
	doi = {10.1016/j.jmb.2018.06.035},
	number = {21},
	urldate = {2024-07-02},
	journal = {Journal of Molecular Biology},
	author = {Stefani, Caroline and Lacy-Hulbert, Adam and Skillman, Thomas},
	month = oct,
	year = {2018},
	pages = {4028--4035},
}

@misc{pidhorskyi_syglass_2018,
	title = {{syGlass}: {Interactive} {Exploration} of {Multidimensional} {Images} {Using} {Virtual} {Reality} {Head}-mounted {Displays}},
	shorttitle = {{syGlass}},
	url = {http://arxiv.org/abs/1804.08197},
	doi = {10.48550/arXiv.1804.08197},
	urldate = {2024-07-02},
	publisher = {arXiv},
	author = {Pidhorskyi, Stanislav and Morehead, Michael and Jones, Quinn and Spirou, George and Doretto, Gianfranco},
	month = aug,
	year = {2018},
	note = {arXiv:1804.08197 [cs]},
}

@article{jaqaman_robust_2008,
	title = {Robust single-particle tracking in live-cell time-lapse sequences},
	volume = {5},
	copyright = {2008 Nature Publishing Group},
	issn = {1548-7105},
	url = {https://www.nature.com/articles/nmeth.1237},
	doi = {10.1038/nmeth.1237},
	language = {en},
	number = {8},
	urldate = {2023-02-20},
	journal = {Nature Methods},
	author = {Jaqaman, Khuloud and Loerke, Dinah and Mettlen, Marcel and Kuwata, Hirotaka and Grinstein, Sergio and Schmid, Sandra L. and Danuser, Gaudenz},
	month = aug,
	year = {2008},
	note = {Number: 8
Publisher: Nature Publishing Group},
	pages = {695--702},
}

@article{elegans_dataset,
	title = {Automated analysis of embryonic gene expression with cellular resolution in {C}. elegans},
	volume = {5},
	issn = {1548-7105},
	doi = {10.1038/nmeth.1228},
	language = {eng},
	number = {8},
	journal = {Nature Methods},
	author = {Murray, John Isaac and Bao, Zhirong and Boyle, Thomas J. and Boeck, Max E. and Mericle, Barbara L. and Nicholas, Thomas J. and Zhao, Zhongying and Sandel, Matthew J. and Waterston, Robert H.},
	month = aug,
	year = {2008},
	pmid = {18587405},
	pmcid = {PMC2553703},
	pages = {703--709},
}

@article{CTC,
	title = {The {Cell} {Tracking} {Challenge}: 10 years of objective benchmarking},
	volume = {20},
	copyright = {2023 The Author(s)},
	issn = {1548-7105},
	shorttitle = {The {Cell} {Tracking} {Challenge}},
	url = {https://www.nature.com/articles/s41592-023-01879-y},
	doi = {10.1038/s41592-023-01879-y},
	language = {en},
	number = {7},
	urldate = {2024-07-03},
	journal = {Nature Methods},
	author = {Maška, Martin and Ulman, Vladimír and Delgado-Rodriguez, Pablo and Gómez-de-Mariscal, Estibaliz and Nečasová, Tereza and Guerrero Peña, Fidel A. and Ren, Tsang Ing and Meyerowitz, Elliot M. and Scherr, Tim and Löffler, Katharina and Mikut, Ralf and Guo, Tianqi and Wang, Yin and Allebach, Jan P. and Bao, Rina and Al-Shakarji, Noor M. and Rahmon, Gani and Toubal, Imad Eddine and Palaniappan, Kannappan and Lux, Filip and Matula, Petr and Sugawara, Ko and Magnusson, Klas E. G. and Aho, Layton and Cohen, Andrew R. and Arbelle, Assaf and Ben-Haim, Tal and Raviv, Tammy Riklin and Isensee, Fabian and Jäger, Paul F. and Maier-Hein, Klaus H. and Zhu, Yanming and Ederra, Cristina and Urbiola, Ainhoa and Meijering, Erik and Cunha, Alexandre and Muñoz-Barrutia, Arrate and Kozubek, Michal and Ortiz-de-Solórzano, Carlos},
	month = jul,
	year = {2023},
	note = {Publisher: Nature Publishing Group},
	pages = {1010--1020},
}

@incollection{jacob_eye_1995,
    title = {Eye {Tracking} in {Advanced} {Interface} {Design}},
    isbn = {978-0-19-507555-7 978-0-19-756031-0},
    url = {https://academic.oup.com/book/41961/chapter/355238840},
    language = {en},
    urldate = {2025-04-23},
    booktitle = {Virtual {Environments} and {Advanced} {Interface} {Design}},
    publisher = {Oxford University Press},
    author = {Jacob, Robert J. K.},
    collaborator = {Jacob, Robert J. K.},
    month = jul,
    year = {1995},
    doi = {10.1093/oso/9780195075557.003.0015},
}

@article{lange_aardvark_2025,
    title = {Aardvark: {Composite} {Visualizations} of {Trees}, {Time}-{Series}, and {Images}},
    volume = {31},
    issn = {1941-0506},
    shorttitle = {Aardvark},
    url = {https://ieeexplore.ieee.org/document/10670500},
    doi = {10.1109/TVCG.2024.3456193},
    number = {1},
    urldate = {2025-06-18},
    journal = {IEEE Transactions on Visualization and Computer Graphics},
    author = {Lange, Devin and Judson-Torres, Robert and Zangle, Thomas A. and Lex, Alexander},
    month = jan,
    year = {2025},
    pages = {1290--1300},
}

@article{pretorius_cell_2015,
    title = {Cell lineage visualisation},
    volume = {34},
    issn = {0167-7055, 1467-8659},
    url = {https://onlinelibrary.wiley.com/doi/10.1111/cgf.12614},
    doi = {10.1111/cgf.12614},
    language = {en},
    number = {3},
    urldate = {2025-06-18},
    journal = {Computer Graphics Forum},
    author = {Pretorius, A. J. and Khan, I. A. and Errington, R. J.},
    month = jun,
    year = {2015},
    pages = {21--30},
}

@article{hong_lineaged_2022,
    title = {{LineageD}: {An} {Interactive} {Visual} {System} for {Plant} {Cell} {Lineage} {Assignments} based on {Correctable} {Machine} {Learning}},
    volume = {41},
    copyright = {© 2022 The Author(s) Computer Graphics Forum © 2022 The Eurographics Association and John Wiley \& Sons Ltd. Published by John Wiley \& Sons Ltd.},
    issn = {1467-8659},
    shorttitle = {{LineageD}},
    url = {https://onlinelibrary.wiley.com/doi/abs/10.1111/cgf.14533},
    doi = {10.1111/cgf.14533},
    language = {en},
    number = {3},
    urldate = {2025-06-18},
    journal = {Computer Graphics Forum},
    author = {Hong, Jiayi and Trubuil, Alain and Isenberg, Tobias},
    year = {2022},
    note = {\_eprint: https://onlinelibrary.wiley.com/doi/pdf/10.1111/cgf.14533},
    pages = {195--207},
}

@article{salvador-martinez_celavi_2021,
    title = {{CeLaVi}: an interactive cell lineage visualization tool},
    volume = {49},
    issn = {0305-1048},
    shorttitle = {{CeLaVi}},
    url = {https://www.ncbi.nlm.nih.gov/pmc/articles/PMC8265160/},
    doi = {10.1093/nar/gkab325},
    number = {W1},
    urldate = {2025-06-19},
    journal = {Nucleic Acids Research},
    author = {Salvador-Martínez, Irepan and Grillo, Marco and Averof, Michalis and Telford, Maximilian J},
    month = may,
    year = {2021},
    pmid = {33956141},
    pmcid = {PMC8265160},
    pages = {W80--W85},
}

@inproceedings{de_leeuw_visualization_2000,
    title = {Visualization of time dependent confocal microscopy data},
    url = {https://ieeexplore.ieee.org/document/885735/},
    doi = {10.1109/VISUAL.2000.885735},
    urldate = {2025-06-19},
    booktitle = {Proceedings {Visualization} 2000. {VIS} 2000 ({Cat}. {No}.{00CH37145})},
    author = {De Leeuw, W.C. and Van Liere, R. and Verschure, P.J. and Visser, A.E. and Manders, E.M.M. and Van Drielf, R.},
    month = oct,
    year = {2000},
    pages = {473--476},
}

@article{guerinot_new_2022,
    title = {New {Approach} to {Accelerated} {Image} {Annotation} by {Leveraging} {Virtual} {Reality} and {Cloud} {Computing}},
    volume = {1},
    issn = {2673-7647},
    doi = {10.3389/fbinf.2021.777101},
    url = {https://www.frontiersin.org/articles/10.3389/fbinf.2021.777101},
    urldate = {2023-11-13},
    journal = {Frontiers in Bioinformatics},
    author = {Guérinot, Corentin and Marcon, Valentin and Godard, Charlotte and Blanc, Thomas and Verdier, Hippolyte and Planchon, Guillaume and Raimondi, Francesca and Boddaert, Nathalie and Alonso, Mariana and Sailor, Kurt and Lledo, Pierre-Marie and Hajj, Bassam and El Beheiry, Mohamed and Masson, Jean-Baptiste},
    year = {2022},
}

@inproceedings{gunther_bionic_2020,
    address = {Cham},
    title = {Bionic {Tracking}: {Using} {Eye} {Tracking} to {Track} {Biological} {Cells} in {Virtual} {Reality}},
    isbn = {978-3-030-66415-2},
    shorttitle = {Bionic {Tracking}},
    doi = {10.1007/978-3-030-66415-2_18},
    language = {en},
    booktitle = {Computer {Vision} – {ECCV} 2020 {Workshops}},
    publisher = {Springer International Publishing},
    author = {Günther, Ulrik and Harrington, Kyle I. S. and Dachselt, Raimund and Sbalzarini, Ivo F.},
    editor = {Bartoli, Adrien and Fusiello, Andrea},
    year = {2020},
    pages = {280--297},
}

@inproceedings{sun_dynamic_2009,
    address = {Richland, SC},
    series = {{AAMAS} '09},
    title = {Dynamic fringe-saving {A}*},
    isbn = {978-0-9817381-7-8},
    doi = {10.5555/1558109.1558136},
    urldate = {2025-06-30},
    booktitle = {Proceedings of {The} 8th {International} {Conference} on {Autonomous} {Agents} and {Multiagent} {Systems} - {Volume} 2},
    publisher = {International Foundation for Autonomous Agents and Multiagent Systems},
    author = {Sun, Xiaoxun and Yeoh, William and Koenig, Sven},
    month = may,
    year = {2009},
    pages = {891--898},
}

@article{pietzsch_bigdataviewer_2015,
    title = {{BigDataViewer}: visualization and processing for large image data sets},
    volume = {12},
    issn = {1548-7091, 1548-7105},
    shorttitle = {{BigDataViewer}},
    url = {https://www.nature.com/articles/nmeth.3392},
    doi = {10.1038/nmeth.3392},
    language = {en},
    number = {6},
    urldate = {2025-06-30},
    journal = {Nature Methods},
    author = {Pietzsch, Tobias and Saalfeld, Stephan and Preibisch, Stephan and Tomancak, Pavel},
    month = jun,
    year = {2015},
    pages = {481--483},
}

@article{platt_micellanngelo_2023,
    title = {{MiCellAnnGELo}: annotate microscopy time series of complex cell surfaces with {3D} virtual reality},
    volume = {39},
    issn = {1367-4811},
    shorttitle = {{MiCellAnnGELo}},
    url = {https://doi.org/10.1093/bioinformatics/btad013},
    doi = {10.1093/bioinformatics/btad013},
    number = {1},
    urldate = {2023-11-13},
    journal = {Bioinformatics},
    author = {Platt, Adam and Lutton, E Josiah and Offord, Edward and Bretschneider, Till},
    month = jan,
    year = {2023},
    pages = {btad013},
}

@article{zhang_4d_2023,
    title = {{4D} {Light}-sheet imaging and interactive analysis of cardiac contractility in zebrafish larvae},
    volume = {7},
    issn = {2473-2877},
    doi = {10.1063/5.0153214},
    number = {2},
    journal = {APL Bioengineering},
    author = {Zhang, X. and Almasian, M. and Hassan, S.S. and Jotheesh, R. and Kadam, V.A. and Polk, A.R. and Saberigarakani, A. and Rahat, A. and Yuan, J. and Lee, J. and Carroll, K. and Ding, Y.},
    year = {2023},
}

@article{amat_fast_2014,
    title = {Fast, accurate reconstruction of cell lineages from large-scale fluorescence microscopy data},
    volume = {11},
    doi = {10.1038/nmeth.3036},
    number = {9},
    journal = {Nature Methods},
    author = {Amat, Fernando and Lemon, William and Mossing, Daniel P and McDole, Katie and Wan, Yinan and Branson, Kristin and Myers, Eugene W and Keller, Philipp J},
    year = {2014},
}
